\documentclass[twocolumn,showpacs,floatfix]{revtex4}

\usepackage{graphics}

\begin{document}

\title{Experimental joint signal-idler quasi-distributions and photon-number statistics for
mesoscopic twin beams}

\author{
Jan Pe\v{r}ina$^{(1,2)}$, Jarom\'\i r K\v{r}epelka$^{(1)}$, Jan
Pe\v{r}ina, Jr.$^{(1)}$, Maria Bondani$^{(3)}$, Alessia
Allevi$^{(4)}$, Alessandra Andreoni$^{(4)}$}

\affiliation{ (1) Joint Laboratory of Optics, Palack\'{y}
University and Institute of Physics of Academy of Sciences of the
Czech Republic, 17. listopadu 50a, 772 07 Olomouc, Czech Republic}
\affiliation{ (2) Department of Optics, Palack\'{y} University,
17. listopadu 50, 772 07 Olomouc, Czech Republic}
\email{perina@prfnw.upol.cz}
\affiliation{ (3) National Laboratory
for Ultrafast and Ultraintense Optical Science C.N.R.-I.N.F.M.,
Via Valleggio 11, 22100 Como, Italy}
\affiliation{ (4) Department
of Physics and Mathematics, University of Insubria and
C.N.R.-I.N.F.M, Via Valleggio 11, 22100 Como, Italy}



\begin{abstract}
Joint signal-idler photoelectron distributions of twin beams
containing several tens of photons per mode have been measured
recently. Exploiting a microscopic quantum theory for joint
quasi-distributions in parametric down-conversion developed
earlier we characterize properties of twin beams in terms of
quasi-distributions using experimental data. Negative values as
well as oscillating behaviour in quantum region are characteristic
for the subsequently determined joint signal-idler
quasi-distributions of integrated intensities. Also the
conditional and difference photon-number distributions are shown
to be sub-Poissonian and sub-shot-noise, respectively.
\end{abstract}

\maketitle

\section{Introduction}

The process of spontaneous parametric down-conversion
\cite{Walls1994,Mandel1995,Perina1994} is one of the fundamental
nonlinear quantum processes that can be understood in terms of
created and annihilated photon pairs. This highly nonclassical
origin of the generated optical fields is responsible for their
unusual properties. They have occurred to be extraordinarily
useful in verification of fundamental laws of quantum mechanics
using tests of Bell inequalities \cite{Perina1994}, generation of
Greenberger-Horne-Zeilinger states \cite{Bouwmeester1999},
demonstration of quantum teleportation \cite{Bouwmeester1997},
quantum cryptography \cite{Lutkenhaus2000}, dense coding, and many
other `quantum protocols' \cite{Bouwmeester2000}. Fields composed
of photon pairs have already found applications, e.g. in quantum
cryptography \cite{Lutkenhaus2000} or metrology \cite{Migdal1999}.
Description of this process has been elaborated from several
points of view for beams containing just one photon pair with a
low probability
\cite{Keller1997,PerinaJr1999,DiGiuseppe1997,Grice1998,PerinaJr2006}
as well as for beams in which many photon pairs occur
\cite{Nagasako2002}. Also stimulated emission of photon pairs has
been investigated
\cite{Lamas-Linares2001,DeMartini2002,Pellicia2003}.

Recent experiments
\cite{Agliati2005,Haderka2005,Haderka2005a,Bondani2007,Paleari2004}
(and references therein) have provided experimental joint
signal-idler photoelectron distributions of twin beams containing
up to several thousands of photon pairs. Extremely sensitive
photodiodes, special single-photon avalanche photodiodes
\cite{Kim1999}, super-conducting bolometers \cite{Miller2003},
time-multiplexed fiber-optics detection loops
\cite{Haderka2004,Rehacek2003,Achilles2004,Fitch2003}, intensified
CCD cameras \cite{Jost1998,Haderka2005}, or methods measuring
attenuated beams \cite{Zambra2005,Zambra2006} are available at
present as detection devices able to resolve photon numbers. Also
homodyne detection has been applied to determine intensity
correlations of twin beams \cite{Vasilyev2000,Zhang2002}. These
advances in experimental techniques have stimulated the
development of a detailed microscopic theory able to describe such
beams and give an insight into their physical properties. A theory
based on multi-mode description of the generated fields has been
elaborated both for spontaneous \cite{Perina2005} as well as
stimulated processes \cite{Perina2006}. This theory allows one to
determine the joint signal-idler quasi-distribution of integrated
intensities from measured joint signal-idler photoelectron
distributions. Considering phases of multi-mode fields generated
in this spontaneous process to be completely random, the joint
signal-idler quasi-distribution of integrated intensities gives us
a complete description of the generated twin beams. As a
consequence of pairwise emission the quasi-distribution of
integrated intensities has a typical shape and attains negative
values in some regions. This quasi-distribution has been already
experimentally reached \cite{Haderka2005,Haderka2005a} for twin
beams containing up to several tens of photon pairs but with mean
numbers of photons per mode being just a fraction of one. Here, we
report on experimental determination of the joint signal-idler
quasi-distribution of integrated intensities for twin beams
containing several tens of photons per mode. Such system may be
considered as mesoscopic and this makes its properties
extraordinarily interesting for an investigation.

Photon-number distributions and quasi-distributions of integrated
intensities provided by theory are contained in Sec.~II. Section~III
is devoted to the analysis of experimental data. Conclusions are drawn in Sec.~IV.

\section{Photon-number distributions and quasi-distributions
of integrated intensities}

\begin{figure}   % fig 1
 \resizebox{0.6\hsize}{!}{\includegraphics{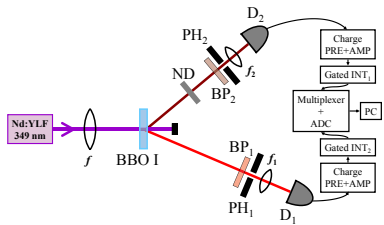}}
 \caption{(Color online) Sketch of the experimental setup. Nd:YLF,
 amplified ps-pulsed laser source; BBO I, nonlinear crystal; $f$,
 $f_{1,2}$,
 lenses; PH$_{1,2}$, 100 $\mu$m diameter pin-holes;
 D$_{1,2}$, pin detectors; BP$_{1,2}$, band-pass filters; ND,
 adjustable neutral-density filter. The boxes on the right-hand
 side indicate the parts of the signal amplification and
 acquisition chains.}
\label{fig1}
\end{figure}

In the experiment, whose layout is sketched in Fig.~\ref{fig1}
(for details, see \cite{Bondani2007}), the third harmonic field
(wavelength 349~nm and time duration 4.45~ps) of an amplified
mode-locked Nd:YLF laser with repetition rate of 500 Hz (High Q
Laser Production, Hohenems, Austria) is used to pump parametric
down-conversion in a BBO crystal (Fujian Castech Crystals, Fuzhou,
China) cut for type-I interaction. The down-converted beams at
wavelengths of 632.8 and 778.2~nm are selected by two 100 $\mu$m
diameter apertures and directed into two amplified pin photodiodes
(S5973-02 and S3883, Hamamatsu Photonics K.K., Japan) using lenses
of appropriate focal lengths (see Fig.~\ref{fig1}). The output
current pulses are digitized and processed by a computer. The
overall detection quantum efficiencies, $ \eta $, are 55~\% for
both arms. First ($ \langle m \rangle $) and second ($ \langle m^2
\rangle $) moments of photoelectron distributions for both signal
and idler beams as well as correlation of photoelectron numbers in
the signal and idler beams are obtained experimentally. Additive
noise present during detection can be measured separately and
subtracted from the measured data. The measured moments of
photoelectron numbers can be corrected also for the overall
quantum efficiency and we then obtain the moments for photons.
Symbol $ \langle n_{1} \rangle $ ($ \langle n_{2} \rangle $) means
mean photon number in signal (idler) field, $ \langle n_{1}^2
\rangle $ ($ \langle n_{2}^2 \rangle $) denotes the second moment
of signal- (idler-) field photon-number distribution, and $
\langle n_{1}n_{2} \rangle $ gives correlations in the number of
signal and idler photons. We note that moments of photon-number
distributions are obtained using the relations:
\begin{eqnarray}   % 1
 \langle n_{i} \rangle &=& \langle m_{i} \rangle / \eta ,
  \nonumber \\
 \langle n_{i}^2 \rangle &=& \langle m_{i}^2 \rangle / \eta^2 -
  (1-\eta)\langle m_i \rangle / \eta^2 , \hspace{5mm}  i=1,2,
  \nonumber \\
 \langle n_1 n_2 \rangle &=& \langle m_1 m_2 \rangle / \eta^2 .
\label{1}
\end{eqnarray}

Moments of integrated intensities can be directly derived from
moments of photon numbers:
\begin{eqnarray}    % 2
 \langle W_i \rangle &=& \langle n_i \rangle , \nonumber \\
 \langle W_i^2 \rangle &=& \langle n_i^2 \rangle - \langle n_i \rangle ,
  \hspace{2mm} i=1,2,  \nonumber \\
 \langle W_1 W_2 \rangle &=& \langle n_1 n_2 \rangle .
\label{2}
\end{eqnarray}
Multi-mode theory of down-conversion developed in
\cite{Perina2005} using a generalized superposition of signal and
noise provides the following relations between the above mentioned
experimental quantities and quantum noise coefficients $ B_1 $, $
B_2 $, $ D_{12} $, and  the number $ M $ of modes:
\begin{eqnarray}    % 3
 \langle W_i \rangle &=& M B_i,  \nonumber \\
 \langle (\Delta W_i)^2 \rangle &=& M B_i^2, \hspace{2mm} i=1,2,
 \nonumber \\
 \langle \Delta W_1 \Delta W_2 \rangle &=& M |D_{12}|^2 .
\label{3}
\end{eqnarray}
The coefficient $ B_i $ gives mean number of photons in mode $ i $
and $ D_{12} $ characterizes mutual correlations between the
signal and idler fields. Inverting relations in Eqs.~(\ref{3}) we
arrive at the expressions for parameters $ B_1 $, $ B_2 $, $ M $,
and $ D_{12} $:
\begin{eqnarray}   % 4
 B_{i} &=& \langle (\Delta W_{i})^2 \rangle/\langle W_{i} \rangle ,
  \nonumber \\
 M_i &=& \langle W_{i} \rangle^2/\langle (\Delta W_{i})^2 \rangle,
  \hspace{2mm} i=1,2 , \nonumber \\
 |D_{12}| &=& \sqrt{ \langle \Delta W_1 \Delta W_2 \rangle/M } .
\label{4}
\end{eqnarray}
As follows from Eqs.~(\ref{4}), the number $ M $ of modes can be
determined from experimental data measured either in the signal or
idler field. This means that the experimental data give two
numbers $ M_1 $ and $ M_2 $ of modes as a consequence of
non-perfect alignment of the setup and non-perfect exclusion of
noise from the data. On the other hand, there occurs only one
number $ M $ of modes (number of degrees of freedom) in the theory
\cite{Perina2005} as it assumes that all pairs of mutually
entangled signal- and idler-field modes are detected at both
detectors. Precise fulfillment of this requirement can hardly be
reached under real experimental conditions. However, experimental
data with $ M_1 \approx M_2 $ can be obtained.

Joint signal-idler photon-number distribution $ p(n_1,n_2) $ for
multi-thermal field with $ M $ degrees of freedom and composed of
photon pairs can be derived in the form \cite{Perina2005}:
\begin{eqnarray}   % 5
 p(n_1,n_2) &=& \frac{1}{\Gamma(M)} \frac{(-K)^{n_2}
 (B_1+K)^{n_1-n_2}}{(1+B_1+B_2+K)^{n_1+M}} \nonumber \\
 & & \hspace{-1cm} \mbox{} \times
  \sum_{r={\rm max}(0,n_2-n_1)}^{n_2} \frac{\Gamma(n_1+M+r)}{r!(n_2-r)!
  (n_1-n_2+r)!} \nonumber \\
 & & \hspace{-1cm} \mbox{} \times  \frac{ [(B_1+K)(B_2+K)]^{r}}{(-K)^r
  (1+B_1+B_2+K)^r}.
\label{5}
\end{eqnarray}
Determinant $ K $ introduced in Eq.~(\ref{5}) and given by the
expression
\begin{equation}    % 6
 K = B_1 B_2 - |D_{12}|^2
\label{6}
\end{equation}
is crucial for the judgement of classicality of a field. Negative
values of the determinant $ K $ mean that a given field cannot be
described classically as in case of the considered field composed
of photon pairs. In Eq.~(\ref{5}), the quantities $ B_1 + K $ and
$ B_2 + K $ cannot be negative and can be considered as
characteristics of fictitious noise present in the signal and
idler fields, respectively. The theory for an ideal lossless case
gives $ K = -B_1 =-B_2 $ together with the joint photon-number
distribution $ p(n_1,n_2) $ in the form of diagonal Mandel-Rice
distribution. On the other hand inclusion of losses and external
noise results in non-diagonal photon-number distribution $
p(n_1,n_2)$ as a consequence of the fact that not all detected
photons are paired. We note that pairing of photons leads to
higher values of elements $ p(n_1,n_2) $ of the joint signal-idler
photon-number distribution around the diagonal $ n_1 = n_2 $ that
violate a classical inequality derived in \cite{Haderka2005}.

A compound Mandel-Rice formula gives the joint signal-idler
photon-number distribution $ p(n_1,n_2) $ at the border between
the classical and nonclassical characters of the field ($ K=0 $):
\begin{equation}   % 7
 p(n_1,n_2)= \frac{ \Gamma(n_1+n_2+M)B_1^{n_1}B_2^{n_2} }{ \Gamma(M)n_1!\, n_2! \,
  (1+B_1+B_2)^{n_1+n_2+M} }.
\label{7}
\end{equation}
If the number $ M $ of modes is large compared to mean values $
\langle n_{1} \rangle $ and $ \langle n_{2} \rangle $ (i.e. for $
B_{1} $, $ B_{2} $, and $ |D_{12}| $ being small) the expression
in Eq.~(\ref{7}) can roughly be approximated by a product of two
Poissonian distributions. If $ K>0 $ or $ K<0 $, weak classical or
quantum fluctuations remain in this Poisson limit of a large
number $ M $ of modes, as follows from the normal generating
function \cite{Perina2005} in the form $ G(\lambda_1,\lambda_2)
\approx \exp(-\lambda_1 \langle n_1 \rangle - \lambda_2 \langle
n_2 \rangle +\lambda_1 \lambda_2 M|D_{12}|^2) $ valid in this
approximation. Thus there are always mode correlations in
agreement with the third formula in Eq.~(\ref{3}).

Declination from an ideal diagonal distribution $ p(n_1,n_2) $
caused by losses can be characterized using conditional
idler-field photon-number distribution $ p_{c,2}(n_2;n_1) $
measured under the condition of detected $ n_1 $ signal photons
and determined along the formula:
\begin{equation}    % 8
 p_{c,2}(n_2;n_1)= p(n_1,n_2)/\sum_{k=0}^{\infty}p(n_1,k) .
\label{8}
\end{equation}
Substitution of Eq.~(\ref{5}) in Eq.~(\ref{8}) leads to the
conditional idler-field photon-number distribution $ p_{c,2} $
with Fano factor $ F_{c,2} $:
\begin{eqnarray}   % 9
 F_{c,2}(n_1) &=& 1 \nonumber \\
 & & \hspace{-2cm} \mbox{} + \frac{(1+M/n_1)[(B_2+K)/(1+B_1)]^2
 -(K/B_1)^2}{ (1+M/n_1)(B_2+K)/(1+B_1)- K/B_1}  \nonumber \\
 & \approx &  1+K/B_1.
\label{9}
\end{eqnarray}
As an approximate expression for the Fano factor $ F_{c,2} $ in
Eq.~(\ref{9}) (valid for $K \approx -B_2$ ) indicates negative
values of the determinant $ K $ are necessary to observe
sub-Poissonian conditional photon-number distributions.
Sub-Poissonian conditional distribution $ p_{c,2} $ emerges from
the formula in Eq.~(\ref{5}) that is a sum of positive terms in
this case. For the ideal lossless case, $ K=-B_1 =-B_2 $ holds and
the Fano factor $ F_{c,2} $ equals 0. On the other hand, positive
values of the determinant $ K $ mean that the sum in Eq.~(\ref{5})
contains large terms with alternating sings (this may lead to
numerical errors in summation) and so the conditional distribution
$ p_{c,2} $ is super-Poissonian. For instance, for $ K $ small
compared to $ B_1 $, $ F_{c,2} = 1+(B_2+K)/(1+B_1) $. We note
that, in this approximation, the value of Fano factor $ F_{c,2} $
equals the value of coefficient $ R $ quantifying sub-shot-noise
correlations and being defined in Eq.~(\ref{11}) below.

Pairing of photons in the detected signal and idler fields leads
to narrowing of distribution $ p_{-} $ of the difference $ n_1 -
n_2 $ of signal- and idler-field photon numbers:
\begin{equation} % 10
 p_{-}(n) = \sum_{n_1,n_2=0}^{\infty} \delta_{n,n_1-n_2} p(n_1,n_2)
\label{10}
\end{equation}
and $ \delta $ denotes Kronecker symbol. If variance of the
difference $ n_1 - n_2 $ of signal- and idler-field photon numbers
is less than the sum of mean photon numbers in the signal and
idler fields we speak about sub-shot-noise correlations and
characterize them by coefficient $ R $ \cite{Bondani2007}:
\begin{equation}   % 11
 R= \frac{ \langle [\Delta(n_1-n_2)]^2 \rangle }{\langle n_1 \rangle +
 \langle n_2 \rangle } <1 .
\label{11}
\end{equation}

Joint signal-idler photon-number distribution $ p(n_1,n_2) $ and
joint signal-idler quasi-distribution $ P_1(W_1,W_2) $ of
integrated intensities belonging to normally-ordered operators are
connected through Mandel's detection equation
\cite{Perina1994,Saleh1978}:
\begin{eqnarray}   % 12
 p(n_1,n_2) &=& \frac{1}{n_1! \, n_2!} \int_{0}^{\infty} dW_1
  \, \int_{0}^{\infty} dW_2 \, W_1^{n_1} W_2^{n_2} \nonumber \\
  & & \mbox{} \times  \exp(-W_1-W_2)
  P_1(W_1,W_2) .
\label{12}
\end{eqnarray}
Relation in Eq.~(\ref{12}) can be generalized to an arbitrary
ordering of field operators \cite{Perina1994,Perina2005} and can
be inverted in terms of series of Laguerre polynomials. Range of
convergence of these series is determined under the condition $ s
\le s_{\rm th} $ where $ s_{th} $ is given in Eq.~(\ref{15})
later. These series define quasi-distributions for $ s
> s_{\rm th} $.

Provided that an $ s $-ordered determinant $ K_s $, $
K_s=B_{1s}B_{2s}-|D_{12}|^2 $ ($ B_{i,s}=B_{i}+(1-s)/2 $, $ i=1,2
$), is positive the $ s $-ordered joint signal-idler
quasi-distribution $ P_s(W_1,W_2) $ of integrated intensities
exists as an ordinary function \cite{Perina2005} which cannot take
on negative values:
\begin{eqnarray}  % 13
 P_s(W_1,W_2) &=&  \frac{1}{\Gamma(M) K_s^M}
  \left(\frac{K_s^2 W_1W_2}{|D_{12}|^2}\right)^{(M-1)/2}
  \nonumber \\
 & & \hspace{0cm} \mbox{} \times \exp\left[-\frac{(B_{2s}W_1/B_{1s}+W_2)B_{1s}}{K_s}\right]
  \nonumber \\
 & & \mbox{} \times  I_{M-1}\left(2
  \sqrt{\frac{|D_{12}|^2W_1W_2}{K_s^2} }\right) .
\label{13}
\end{eqnarray}
Symbol $ I_{M} $ denotes modified Bessel function and $ \Gamma $
stands for $ \Gamma $-function.

If the $ s $-ordered determinant $ K_s $ is negative, the joint
signal-idler quasi-distribution $ P_s $ of integrated intensities
exists in general as a generalized function that can be negative
or even have singularities. It can be approximated by the
following formula \cite{Perina2005}:
\begin{eqnarray}   % 14
 P_s(W_1,W_2) & \approx & \frac{A(W_1W_2)^{(M-1)/2}}{\pi
\Gamma(M) (B_{1s}B_{2s})^{M/2}}
  \exp\left(-\frac{W_1}{2B_{1s}}-\frac{W_2}{2B_{2s}}\right) \nonumber \\
 & & \mbox{} \times
  \rm{sinc}\left[A\left(\frac{B_{2s}}{B_{1s}}W_1-W_2\right)\right];
\label{14}
\end{eqnarray}
$ {\rm sinc}(x) = \sin(x)/x $. Oscillating behaviour is typical
for the quasi-distribution $ P_s $ written in Eq.~(\ref{14}).

There exists a threshold value $ s_{\rm th} $ of the ordering
parameter $ s $ for given values of parameters $ B_1 $, $ B_2 $,
and $ D_{12} $ determined by the condition $ K_s=0 $:
\begin{equation}  % 15
 s_{\rm th} = 1 + B_1+B_2 - \sqrt{(B_1+B_2)^2 -
4 K}; \label{15}
\end{equation}
$ -1 \le s_{\rm th} \le 1 $. Quasi-distributions $ P_s $ for $ s\le
s_{\rm th} $ are ordinary functions with non-negative values whereas
those for $ s > s_{\rm th} $ are generalized functions with negative
values and oscillations.

Similarly as for photon numbers we can define quasi-distribution $
P_{s,-} $ of the difference $ W_1-W_2 $ of signal- and idler-field
integrated intensities as a quantity useful for description of
photon pairing:
\begin{eqnarray}    % 16
 P_{s,-}(W) &=& \int_{0}^{\infty} dW_1 \, \int_{0}^{\infty} dW_2 \,
  \nonumber \\
 & & \mbox{} \times \delta(W - W_1 + W_2) P_s(W_1,W_2) .
\label{16}
\end{eqnarray}
Quasi-distribution $ P_{s,-} $ oscillates and takes on negative
values as a consequence of pairwise character of the detected
fields if $s \ge s_{\rm th}$.

There exists relation between variances of the difference $ n_1 -
n_2 $ of signal- and idler-field photon numbers and difference $
W_1 - W_2 $ of signal- and idler-field integrated intensities:
\begin{equation}   % 17
 \langle [\Delta(n_1-n_2)]^2 \rangle = \langle n_1 \rangle +
\langle n_2 \rangle + \langle [\Delta(W_1-W_2)]^2  \rangle .
\label{17}
\end{equation}
According to Eq.~(\ref{17}) negative values of the
quasi-distribution $ P_{s,-} $ (as well as these of
quasi-distribution $ P_s $) are necessary to observe
sub-shot-noise correlations in signal- and idler-field photon
numbers as described by the condition $ R < 1 $ in Eq.~(\ref{11}).

\section{Experimental distributions}

As an example, we analyze the following experimental data
appropriate for photons and derived from the experimental data for
photoelectrons given in \cite{Bondani2007} using relations in
Eqs.~(\ref{1}) [$ \eta=0.55 $]:
\begin{eqnarray}  % 18
 \langle n_1 \rangle = 959.21 , & & \langle n_2 \rangle = 1078.3 ,
  \nonumber \\
 \langle n_1^2 \rangle = 971829.7 , & & \langle n_2^2 \rangle =
  1218608 , \nonumber \\
 \langle n_1 n_2 \rangle = 1088083 . & &
\end{eqnarray}
These data thus characterize photon fields, as they have been
obtained after correction for the nonunit detection efficiency.
Formulas in Eqs.~(\ref{2}) and (\ref{4}) then give mean number $
B_1 $ of signal photons per mode, mean number $ B_2 $ of idler
photons per mode, number $ M_1 $ of signal-field modes, and number
$ M_2 $ of idler-field modes:
\begin{eqnarray}   % 19
 B_1=52.95, & & B_2=50.81 , \nonumber \\
 M_1=18.11, & & M_2=21.22.
\label{19}
\end{eqnarray}
Numbers $ M_1 $ and $ M_2 $ of modes given in Eqs.~(\ref{19}) and
determined from data characterizing signal ($ M_1 $) and idler ($
M_2 $) fields slightly differ owing to experimental imperfections.
That is why we use a mean number $ M $ of modes [$ M = (M_1 +
M_2)/2 $] and determine the coefficient $ D_{12} $ along the
relation in Eqs.~(\ref{4}):
\begin{equation}  % 20
 M=19.66, \hspace{5mm} |D_{12}|=52.29 .
\label{20}
\end{equation}
Determinant $ K $ given in Eq.~(\ref{6}) then equals -44.23, i.e.
the measured field is nonclassical. Coefficient $ R $ defined in
Eq.~(\ref{11}) equals 0.19 (-7.2 dB reduction of vacuum
fluctuations) and this means that fluctuations in the difference $
n_1 - n_2 $ of signal and idler photon numbers are below
shot-noise level. This also means [see Eq.~(\ref{17})] that
variance $ \langle [\Delta (W_1-W_2)]^2 \rangle $ of the
difference of signal- and idler-field integrated intensities is
negative ($ \langle [\Delta(W_1-W_2)]^2 \rangle = - 1654 $).
Negative value of this variance is caused by pairwise character of
the detected fields, which leads to strong correlations in
integrated intensities $ W_1 $ and $ W_2 $. Also the value of
covariance $ C $ ($ C=\langle \Delta n_1 \Delta n_2 \rangle/
\sqrt{ \langle [\Delta n_1]^2\rangle \langle[\Delta n_2]^2 \rangle
} $) of signal $ n_1 $ and idler $ n_2 $ photon numbers close to
one ($ C=0.997 $) is evidence of a strong pairwise character of
the detected fields. We note that also a two-mode principal
squeeze variance $ \lambda $ characterizing phase squeezing and
related to one pair of modes can be determined along the formula:
\begin{equation}   % 21
 \lambda = 1 + B_1 + B_2 - 2|D_{12}|.
\label{21}
\end{equation}
We arrive at $ \lambda = 0.18 $ using our data in Eq.~(\ref{21})
and so the generated field is also phase squeezed.

\begin{figure}[tbp]   % fig. 2
 \resizebox{0.8\hsize}{!}{\includegraphics{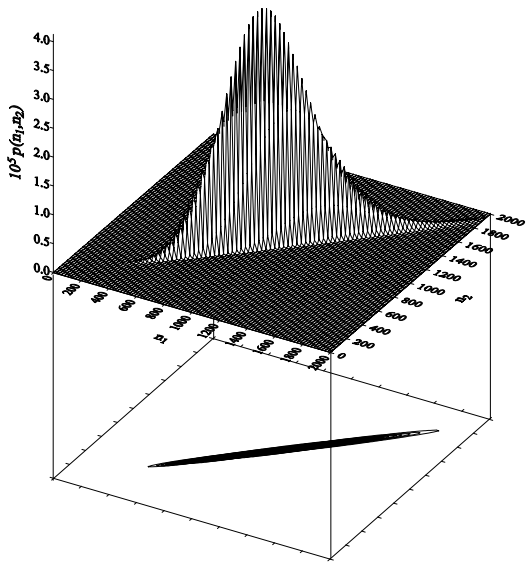}}
 \caption{Joint signal-idler photon-number distribution $ p(n_1,n_2)
  $.}
\label{fig2}
\end{figure}

The joint signal-idler photon-number distribution $ p(n_1,n_2) $
determined along the formula in Eq.~(\ref{5}) for values of
parameters in Eqs.~(\ref{19}) and (\ref{20}) is shown in
Fig.~\ref{fig2}. Strong correlations in signal-field $ n_1 $ and
idler-field $ n_2 $ photon numbers are clearly visible. Nonzero
elements of the joint photon-number distribution $ p(n_1,n_2) $
are localized around a line given by the condition $ n_1 \approx
n_2 $ as documented in contour plot in Fig.~\ref{fig2}.

\begin{figure}[tbp]   % fig. 3
 \resizebox{0.6\hsize}{!}{\includegraphics{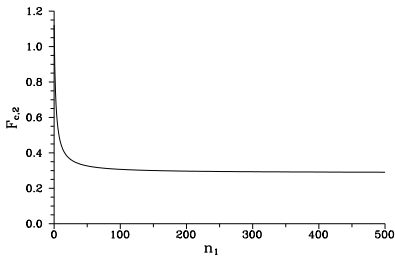}}
 \caption{Fano factor $ F_{c,2} $ of the conditional signal-idler photon-number
  distribution $ p_{c,2} $ as a function of the number $ n_1 $ of detected signal photons.}
\label{fig3}
\end{figure}

Conditional distributions $ p_{c,2} $ of idler-field photon
numbers $ n_2 $ conditioned by detection of a given number $ n_1 $
of signal photons defined in Eq.~(\ref{8}) are also sub-Poissonian
(see Fig.~\ref{fig3}). The greater the value of the number $ n_1 $
of signal photons the smaller the value of Fano factor $ F_{c,2} $
given in Eq.~(\ref{9}). If mean numbers $ \langle n_1 \rangle $
and $ \langle n_2 \rangle $ of signal- and idler-field photons are
small compared to the number $ M $ of modes the joint
photon-number distribution $ p(n_1,n_2) $ behaves like a product
of two Poissonian distributions and so $ F_{c,2} \approx 1 $. Fano
factor $ F_{c,2} $ reaches its asymptotic value after certain
value of the number $ n_1 $ of signal-field photons [see
discussion below Eq.~(\ref{7})].

\begin{figure}[tbp]   % fig. 4
 \resizebox{0.6\hsize}{!}{\includegraphics{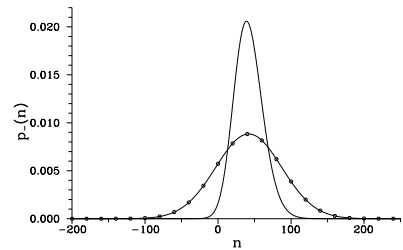}}
 \caption{Distributions $ p_{-}(n) $ of the difference $ n $ of
  signal-field ($ n_1 $) and idler-field ($ n_2 $) photon numbers; $ n=n_1-n_2 $.
  Solid curve without symbols characterizes the experimental data. Solid curve with
  $ \circ $ gives the distribution obtained from the joint signal-idler photon-number
  distribution in the form of product of two independent Poissonian
  distributions with mean photon-numbers given by
  experimental data and is shown for comparison.}
\label{fig4}
\end{figure}

Strong correlations in signal-field $ n_1 $ and idler-field $ n_2
$ photon numbers lead to sub-Poissonian distribution $ p_{-} $ of
the difference $ n_1 - n_2 $ of photon numbers defined in
Eq.~(\ref{10}) (see Fig.~\ref{fig4}).

\begin{figure}[tbp]  % figs. 5a,b
 {\raisebox{6.5 cm}{a)} \hspace{5mm}
 \resizebox{0.8\hsize}{!}{\includegraphics{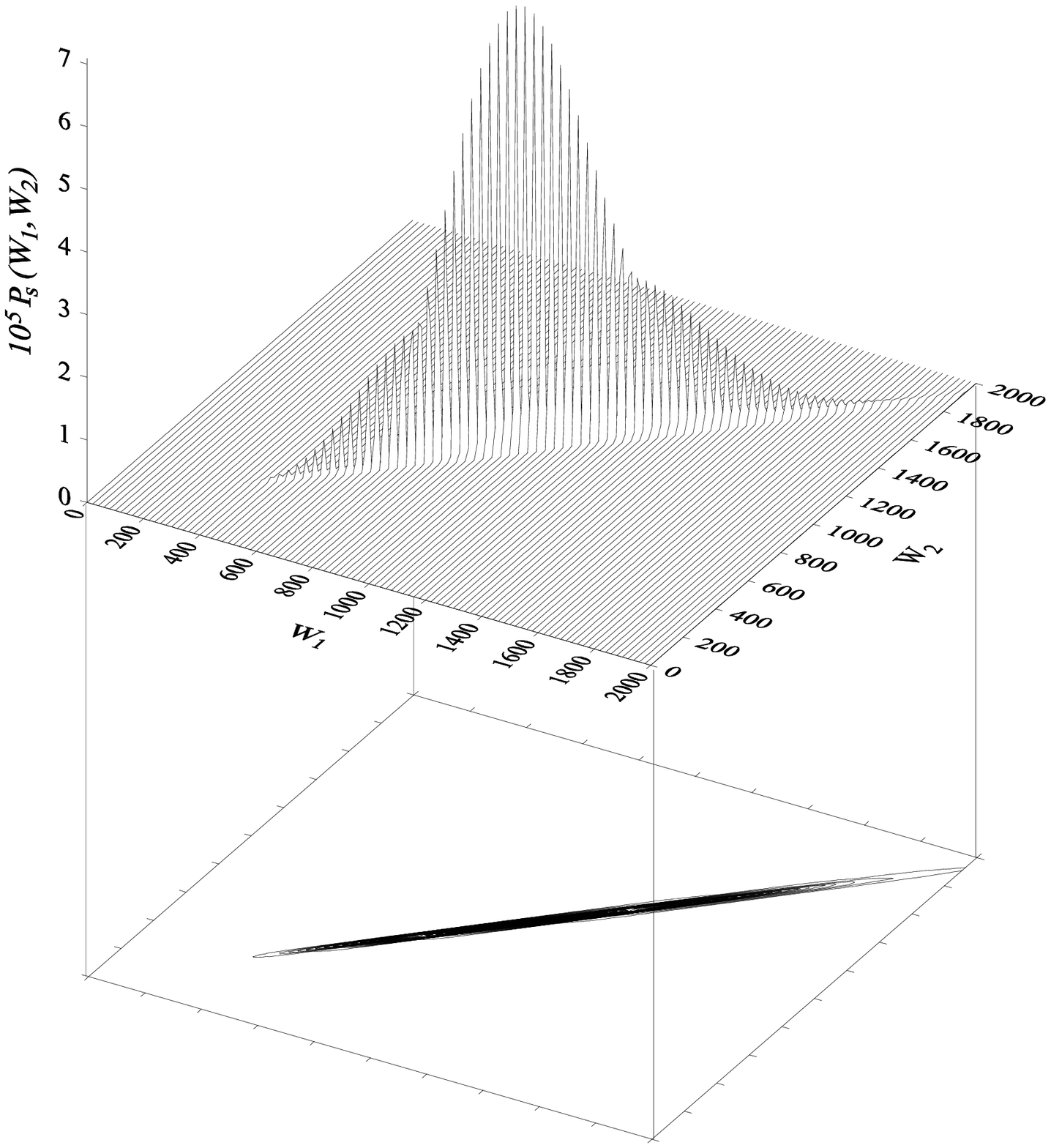}}

 \vspace{5mm}
 \raisebox{6.5 cm}{b)} \hspace{5mm}
 \resizebox{0.8\hsize}{!}{\includegraphics{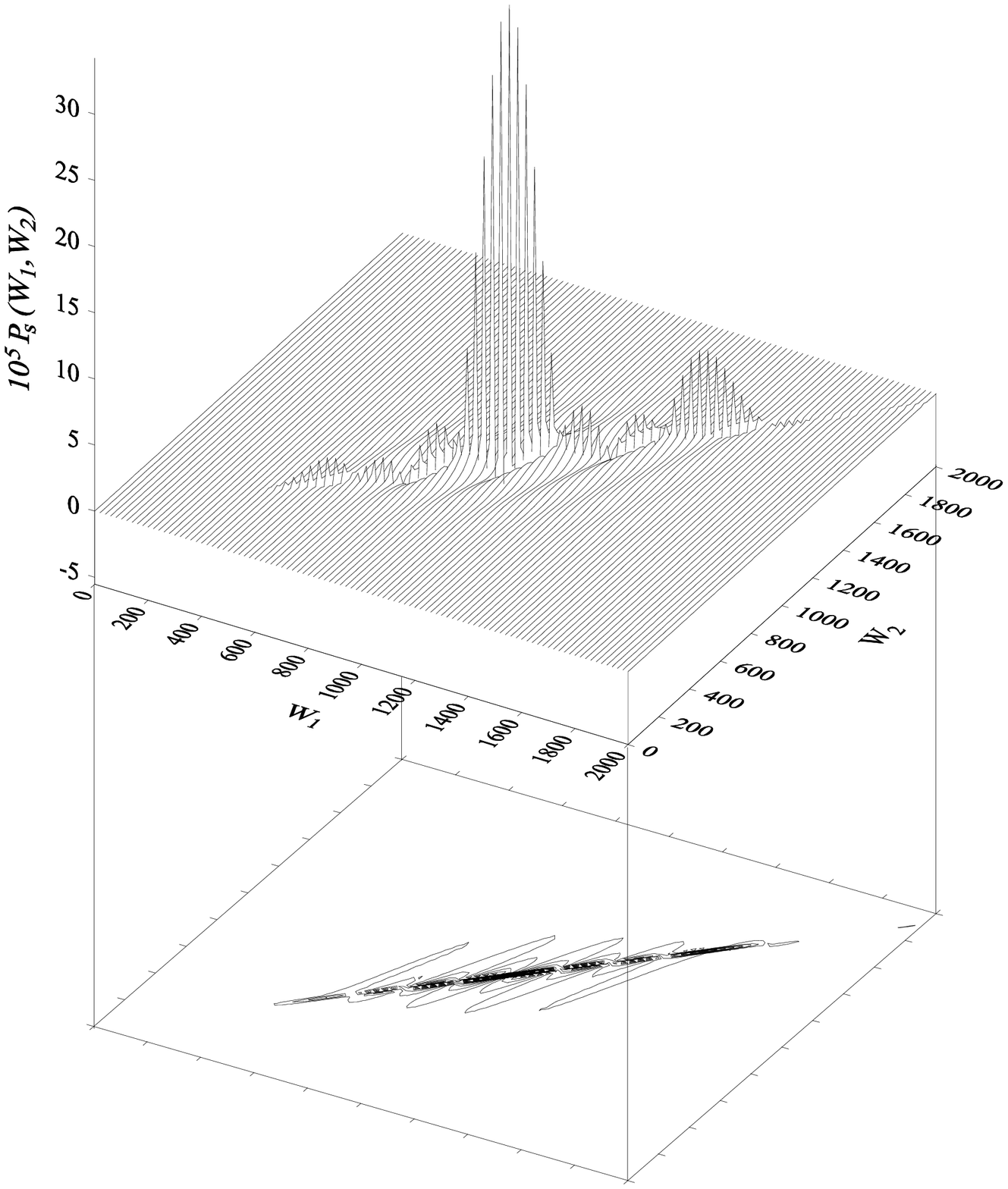}}}
 \vspace{0mm}

 \caption{Joint signal-idler quasi-distributions $ P_s(W_1,W_2) $ of integrated
 intensities of signal ($ W_1 $) and idler ($ W_2 $) fields for $ s=0.1 $
 (a) and $ s=0.2 $ (b).}
\label{fig5}
\end{figure}

Joint signal-idler quasi-distributions $ P_s(W_1,W_2) $ of
integrated intensities differ qualitatively according to the value
of ordering parameter $ s $ ($ s_{\rm th} = 0.15 $ for the
experimental data). Nonclassical character of the detected fields
is smoothed out ($ K_s=2.66>0 $) for the value of $ s $ equal to
0.1 as shown in Fig.~\ref{fig5}(a). On the other hand, the value
of $ s $ equal to 0.2 is sufficient to observe quantum features ($
K_s=-2.53<0 $) in the joint signal-idler quasi-distribution $
P_s(W_1,W_2) $ that is plotted in Fig.~\ref{fig5}(b). In this case
oscillations and negative values occur in the graph of the joint
quasi-distribution $ P_s(W_1,W_2) $.

Determination of the number $ M $ of modes for the overall field
has to be done carefully because it might happen that the theory
shows nonphysical results. There are three conditions determining
the region with nonclassical behavior: $ K<0 $, $ K+B_{1}>0 $, and
$ K+B_{2}>0 $. These conditions can be transformed into the
following inequalities:
\begin{equation}  % 22
 B_1 B_2 < |D_{12}|^2 < B_1 B_2+\min(B_1,B_2) ,
\end{equation}
where $ \min $ denotes minimum value of its arguments. If
sub-shot-noise reduction in fluctuations of the difference $ n_1 -
n_2 $ of signal- and idler-field photon numbers is assumed
(implying $ \langle[\Delta(W_1-W_2)]^2\rangle<0 $), even stronger
conditions can be derived:
\begin{equation}   % 23
 B_1 B_2 \leq (B_1^2+B_2^2)/2 < |D_{12}|^2 < B_1 B_2+\min(B_1,B_2)
\end{equation}
and we therefore need to fulfill the inequality: $ (B_1-B_2)^2
\leq 2 \min(B_1,B_2) $. Assuming $B_1 \geq B_2$ we arrive at the
final condition:
\begin{equation}   % 24
 B_1 \leq B_2+\sqrt{2B_2}.
\label{24}
\end{equation}
The condition in Eq.~(\ref{24}) gives limitation to the lowest
possible physical value of the number $ M $ of modes. Increasing
the value of number $ M $ of modes from this boundary value the
field behaves nonclassically first and then its properties become
classical.

Nonclassical character of the detected field is given by the
condition $ K < 0 $ in theory. In experiment we usually measure
coefficient $ R $ given in Eq.~(\ref{11}) in order to prove
nonclassical character of the field given by the condition $ R < 1
$. According to the developed theory \cite{Perina2005}, if the
field is classical ($ K > 0 $), then there is no sub-shot-noise
reduction in fluctuations of the difference of signal- and
idler-field photon numbers ($ R
> 1 $). On the other hand, situation is more complicated for
nonclassical fields with $ K < 0 $. Provided that $ B_1=B_2=B $,
negative value of the determinant $ K $ implies $ B^2-|D_{12}|^2<0
$ and $ \langle[\Delta(W_1-W_2)]^2\rangle = 2M(B^2-|D_{12}|^2)<0
$. Thus the use of relation in Eq.~(\ref{17}) gives $ R < 1 $,
i.e. we have sub-shot-noise reduction of fluctuations in the
difference of photon numbers. If $ B_1 \neq B_2 $, it may happen
that $ R \ge 1 $, i.e. non-classicality of the field is not
observed in sub-shot-noise reduction of fluctuations of the
difference of photon numbers. We note that even conditional
photon-number distributions $ p_{c,2} $ can remain sub-Poissonian
in this case.

The above discussion has been devoted to statistical properties of
photons. Qualitatively similar results can be obtained also for
photoelectrons. Quantum Burgess theorem assures that
sub-Poissonian photoelectron distribution occurs provided that
photon-number distribution is also sub-Poissonian [$ F_m -1=\eta
(F_n-1) $, $ F_n $ ($ F_m $) means Fano factor for photons
(photoelectrons), $ \eta $ is detection efficiency]. Photoelectron
distributions are noisier compared to photon-number distributions
and that is why nonclassical properties of photoelectron
distributions are weaker.

\section{Conclusions}

Nonclassical character of mesoscopic twin beams containing several
tens of photon pairs per mode has been demonstrated using
experimental data. Joint signal-idler photon-number distribution,
its conditional photon-number distributions, distribution of the
difference of signal- and idler-field photon numbers, and joint
signal-idler quasi-distributions of integrated intensities have
been determined to provide evidence of non-classicality of the
detected twin beams.

\acknowledgments This work was supported by projects KAN301370701
of Grant agency of AS CR, 1M06002 and MSM6198959213 of the Czech
Ministry of Education, and FIRB-RBAU014CLC-002 of the Italian
Ministry of University and Scientific Research.


\begin{thebibliography}{99}

\bibitem{Walls1994} D.F. Walls and G.J. Milburn, {\it Quantum Optics}
 (Springer, Berlin, 1994) chap. 5.
\bibitem{Mandel1995} L. Mandel and E. Wolf, {\it Optical Coherence and Quantum
 Optics} (Cambridge Univ. Press, Cambridge, 1995) chap. 22.4.
\bibitem{Perina1994} J. Pe\v{r}ina, Z. Hradil, and B. Jur\v{c}o, {\it Quantum
 Optics and Fundamentals of Physics} (Kluwer, Dordrecht, 1994) chap. 8.
\bibitem{Bouwmeester1999} D. Bouwmeester, J.-W. Pan, M. Daniell,
 H. Weinfurter, and A. Zeilinger, Phys. Rev. Lett. {\bf 82},
 1345 (1999).
\bibitem{Bouwmeester1997} D. Bouwmeester, J.W. Pan,
 K. Mattle, M. Eibl, H. Weinfurter, and A. Zeilinger,
 Nature {\bf 390}, 575 (1997).
\bibitem{Lutkenhaus2000} D. Bru\ss{}, N. L\"{u}tkenhaus,
 in {\it Applicable Algebra in Engineering, Communication
 and Computing} Vol. 10 (Springer, Berlin, 2000); p. 383.
\bibitem{Bouwmeester2000} D. Bouwmeester, A. Ekert, and A. Zeilinger
 (Eds.) {\it The Physics of Quantum Information} (Springer, Berlin,
 2000).
\bibitem{Migdal1999} A. Migdall, Physics Today {\bf 1}, 41 (1999).
\bibitem{Keller1997} T.E. Keller and M.H. Rubin, Phys. Rev. A
{\bf 56}, 1534 (1997).
\bibitem{PerinaJr1999} J. Pe\v{r}ina, Jr., A.V. Sergienko, B.M. Jost,
 B.E.A. Saleh, and M.C. Teich, Phys. Rev. A {\bf 59},
 2359 (1999).
\bibitem{DiGiuseppe1997} G. Di Giuseppe, L. Haiberger, F. De
 Martini, and A.V. Sergienko, Phys. Rev. A {\bf 56}, R21 (1997).
\bibitem{Grice1998} W.P. Grice, R. Erdmann, I.A. Walmsley, and
 D. Branning, Phys. Rev. A {\bf 57}, R2289 (1998).
\bibitem{PerinaJr2006} J. Pe\v{r}ina, Jr., M. Centini, C. Sibilia, M. Bertolotti,
 and M. Scalora, Phys. Rev. A {\bf 73}, 033823 (2006); arXiv:quant-ph/0604017.
\bibitem{Nagasako2002} E.M. Nagasako, S.J. Bentley, R.W. Boyd, and G.S.
 Agarwal, J. Mod. Opt. {\bf 49}, 529 (2002).
\bibitem{Lamas-Linares2001} A. Lamas-Linares, J.C. Howell, and D. Bouwmeester,
 Nature {\bf 412}, 887 (2001).
\bibitem{DeMartini2002}  F. De Martini, V. Bu\v{z}ek, F.
 Sciarrino, and C. Sias, Nature {\bf 419}, 815 (2002).
\bibitem{Pellicia2003} D. Pelliccia, V. Schettini, F. Sciarrino, C. Sias, and F. De
 Martini, Phys. Rev. A {\bf 68}, 042306 (2003).
\bibitem{Agliati2005} A. Agliati, M. Bondani, A. Andreoni, G. De Cillis, and M.G.A. Paris,
 J. Opt. B: Quant. Semiclass. Opt. {\bf 7}, S652 (2005).
\bibitem{Haderka2005} O. Haderka, J. Pe\v{r}ina, Jr., M. Hamar, and J.
 Pe\v{r}ina, Phys. Rev. A {\bf 71}, 033815 (2005).
\bibitem{Haderka2005a} O. Haderka, J. Pe\v{r}ina, Jr., M. Hamar, J. Opt. B:
 Quantum Semiclass. Opt. {\bf 7}, S572 (2005).
\bibitem{Bondani2007} M. Bondani, A. Allevi, G. Zambra, M.G.A. Paris, and A.
 Andreoni, Phys. Rev. A {\bf 76}, 013833 (2007); arXiv:quant-ph/0612198v1.
\bibitem{Paleari2004} F. Paleari, A. Andreoni, G. Zambra, and M. Bondani,
 Opt. Express {\bf 12}, 2816 (2004).
\bibitem{Kim1999} J.~Kim, S.~Takeuchi, Y.~Yamamoto, and H.~H.~Hogue,
 Appl. Phys. Lett. {\bf 74}, 902 (1999).
\bibitem{Miller2003} A.J. Miller, S.W. Nam, J.M. Martinis, and A.V.
 Sergienko, Appl. Phys. Lett. {\bf 83}, 791 (2003).
\bibitem{Haderka2004} O. Haderka, M. Hamar, and J. Pe\v{r}ina, Jr., Eur. Phys. J. D 28,
 149 (2004); arXiv:quant-ph/0302154.
\bibitem{Rehacek2003} J. \v{R}eh\'{a}\v{c}ek, Z. Hradil, O. Haderka,
 J. Pe\v{r}ina, Jr., and M. Hamar,
 Phys Rev. A {\bf 67}, 061801(R) (2003); arXiv:quant-ph/0303032.
\bibitem{Achilles2004} D. Achilles, Ch. Silberhorn, C. Sliwa, K.
 Banaszek, and I.A. Walmsley, J. Mod. Opt. {\bf 51}, 1499 (2004);
 arXiv:quant-ph/0305191.
\bibitem{Fitch2003} M.J. Fitch, B.C. Jacobs, T.B. Pittman, and J.D.
 Franson, Phys. Rev. A {\bf 68}, 043814 (2003);
 arXiv:quant-ph/0305193.
\bibitem{Jost1998} B.M. Jost, A.V. Sergienko, A.F. Abouraddy,
 B.E.A. Saleh, and M.C. Teich, Opt. Express {\bf 3}, 81 (1998).
\bibitem{Zambra2005} G. Zambra, A. Andreoni, M. Bondani, M. Gramegna,
 M. Genovese, G. Brida, A. Rossi, and M.G.A. Paris, Phys. Rev. Lett.
{\bf 95}, 063602 (2005).
\bibitem{Zambra2006} G. Zambra and M.G.A. Paris, Phys. Rev. A
{\bf 74}, 063830 (2006).
\bibitem{Vasilyev2000} M. Vasilyev, S.-K. Choi, P. Kumar, and
 G.M. D'Ariano, Phys. Rev. Lett. {\bf 84}, 2354 (2000).
\bibitem{Zhang2002} Y. Zhang, K. Kasai, and M. Watanabe, Opt. Lett. {\bf 27}, 1244
 (2002).
\bibitem{Perina2005} J. Pe\v{r}ina and J. K\v{r}epelka, J. Opt. B: Quant.
 Semiclass. Opt. {\bf 7}, 246 (2005).
\bibitem{Perina2006} J. Pe\v{r}ina and J. K\v{r}epelka, Opt. Commun. {\bf
 265}, 632 (2006).
\bibitem{Saleh1978} B.E.A. Saleh, Photoelectron Statistics (Springer-Verlag, New York,
 1978).

\end{thebibliography}
\end{document}